\documentclass[letterpaper]{article} 
\usepackage{aaai25}  
\usepackage{times}  
\usepackage{helvet}  
\usepackage{courier}  
\usepackage[hyphens]{url}  
\usepackage{graphicx} 
\usepackage{natbib}  
\usepackage{caption} 
\usepackage{algorithm}
\usepackage{algorithmic}

\usepackage{xspace}
\usepackage{makecell}
\usepackage{multirow}
\usepackage[table,xcdraw]{xcolor}
\usepackage{subcaption}
\usepackage{graphicx}
\usepackage[justification=centering]{caption}
\usepackage{pdfpages}
\usepackage{booktabs}
\usepackage{soul, color}
\usepackage{url}

\usepackage{newfloat}
\usepackage{listings}
\DeclareCaptionStyle{ruled}{labelfont=normalfont,labelsep=colon,strut=off} 
\floatstyle{ruled}
\newfloat{listing}{tb}{lst}{}
\floatname{listing}{Listing}
\usepackage{newfloat}
\usepackage{listings}
\DeclareCaptionStyle{ruled}{labelfont=normalfont,labelsep=colon,strut=off} 
\floatstyle{ruled}
\newfloat{listing}{tb}{lst}{}
\floatname{listing}{Listing}
\newcommand*{\eg}{e.g.,\xspace}
\newcommand*{\ie}{i.e.,\xspace}

\title{Making Transparency Advocates: An Educational \\ Approach Towards Better Algorithmic Transparency in Practice}

\author{
    Andrew Bell,
    Julia Stoyanovich
}
\affiliations{
    New York University, New York, NY, USA\\
    alb9742@nyu.edu, stoyanovich@nyu.edu
}

\begin{document}
\maketitle

\begin{abstract}
Concerns about the risks and harms posed by artificial intelligence (AI) have resulted in significant study into algorithmic transparency, giving rise to a sub-field known as Explainable AI (XAI). Unfortunately, despite a decade of development in XAI, an existential challenge remains: progress in research has not been fully translated into the actual implementation of algorithmic transparency by organizations. In this work, we test an approach for addressing the challenge by creating transparency advocates, or motivated individuals within organizations who drive a ground-up cultural shift towards improved algorithmic transparency.

Over several years, we created an open-source educational workshop on algorithmic transparency and advocacy. We delivered the workshop to professionals across two separate domains to improve their algorithmic transparency literacy and willingness to advocate for change. In the weeks following the workshop, participants applied what they learned, such as speaking up for algorithmic transparency at an organization-wide AI strategy meeting. We also make two broader observations: first, advocacy is not a monolith and can be broken down into different levels. Second, individuals' willingness for advocacy is affected by their professional field. For example, news and media professionals may be more likely to advocate for algorithmic transparency than those working at technology start-ups.
\end{abstract}

\section{Introduction}
\label{sec:introduction}

\begin{figure*}
     \centering
     \begin{subfigure}[b]{0.22\textwidth}
         \centering
         \includegraphics[width=\textwidth]{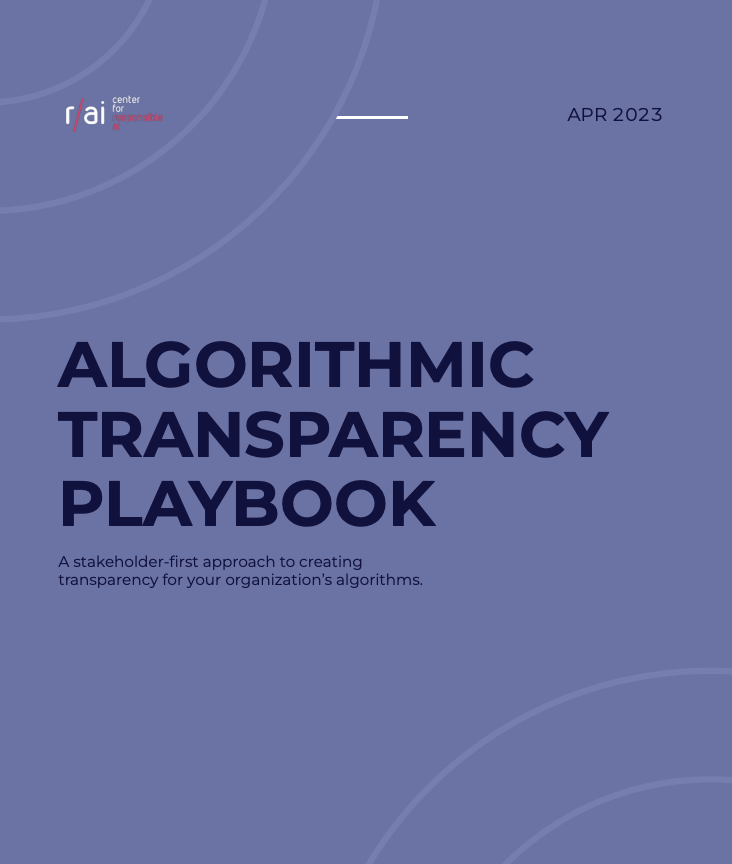}
         \caption{Cover of the \emph{Algorithmic Transparency Playbook}, available for download at \url{https://r-ai.co/algorithmic-transparency-playbook}.}
         \label{fig:y equals x}
     \end{subfigure}
     \hfill
     \begin{subfigure}[b]{0.74\textwidth}
         \centering
         \includegraphics[width=\textwidth]{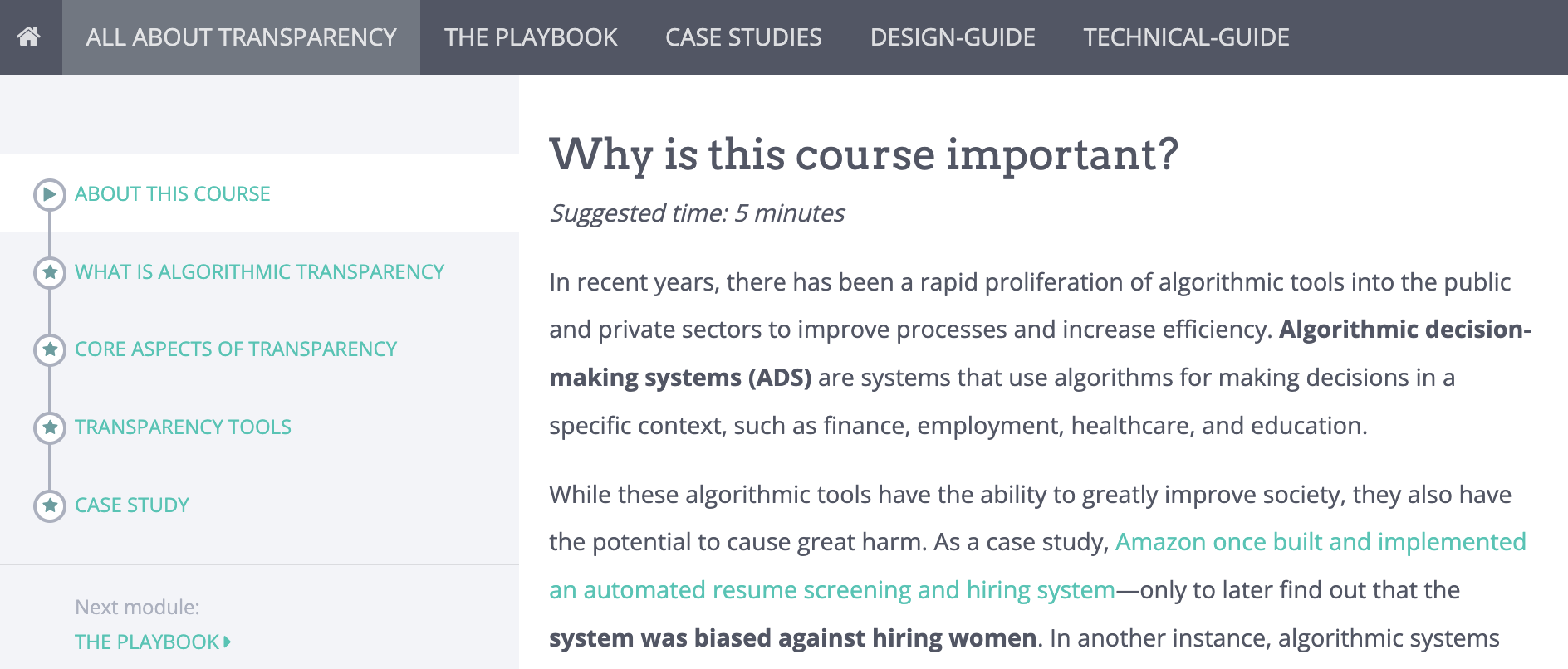}
         \caption{The free open-source online course is available at \url{https://r-ai.co/transparency-playbook-course}.}
         \label{fig:three sin x}
     \end{subfigure}
\end{figure*}

There are widespread concerns about the significant risks posed by artificial intelligence (AI) systems in both the public and private sectors, particularly for marginalized or historically disadvantaged groups~\cite{hu_2020, Sapiezynski2017AcademicPP, obermeyer2019dissecting}. One major risk factor, compounded by the release of Large Language Models, is the lack of transparency in AI systems that make high-stakes decisions~\cite{rudin2019stop,kirilenko2017flash}. These concerns have led to the emergence of  \emph{Explainable Artificial Intelligence} (XAI), a sub-field focused on studying how well AI systems can be understood by humans~\cite{bell2023algorithmic}. While significant progress has been made in developing and evaluating methods for explaining complex AI systems---through multi-disciplinary approaches combining machine learning and human-computer interaction~\cite{DBLP:conf/nips/LundbergL17, ribeiro2016should, datta2016algorithmic, DBLP:journals/corr/abs-2004-00668,abdul2020cogam, yang2019study, holzinger2020measuring, bell2022s}---evidence suggests that companies and organizations using AI often undervalue or remain unaware of these methods~\cite{dastin2022amazon, hill2022secretive}. As a result, XAI faces an existential challenge: how can we move beyond the research setting to \emph{ensure the real-world implementation of transparent AI systems}~\cite{beattie2022challenges}?

While government regulation seems like the natural solution to this challenge, the rapid development of AI technologies has greatly outpaced public oversight, resulting in an incomplete patchwork of laws and regulations~\cite{jobin2019artifical}. To date, over 70 nations and intergovernmental organizations have published over 1,000 AI strategies, actions plans, policy papers, or directives~\cite{oecd}. Unfortunately, many of these efforts face a significant limitation: they remain uncertain about how to meaningfully implement transparency~\cite{jobin2019artifical, loi2021towards, DBLP:journals/internet/GasserA17}. For example, in the United States, the Biden Administration has issued broad AI guidance under the \emph{Executive Order on Safe, Secure, and Trustworthy Development and Use of Artificial Intelligence}~\cite{whitehouse}, but the U.S. Congress has taken little action to strengthen regulations or enact specific laws governing AI transparency practices.

Meanwhile, the private sector demonstrates inconsistent interest in algorithmic transparency and responsible AI practices. As an example, during substantial layoffs in May 2023, Microsoft disbanded its entire AI ethics team.\footnote{\url{https://www.theverge.com/2023/3/13/23638823/microsoft-ethics-society-team-responsible-ai-layoffs}} In light of these challenges, this work explores a complementary pathway to ensuring safe, transparent AI: educating and empowering \emph{transparency advocates} within organizations.

We define transparency advocates as a subset of what~\citet{meyerson2003tempered} called ``tempered radicals,'' or committed employees who drive institutional change over time (sometimes clandestinely), with a focus on algorithmic transparency. Tempered radicals can be very effective: in one example, over a 30-year period, a Black senior executive at a large West Coast bank covertly hired 3,500 women and minority employees to improve company diversity. We hypothesize that transparency advocates can similarly drive significant, bottom-up organizational change toward improved algorithmic transparency.

\paragraph*{Study approach.} Over several years, we created a workshop on algorithmic transparency that provides an overview, introduces best practices and tools for implementing transparency, and outlines strategies for advocating it. This workshop is part of the education and training mission of the Center for Responsible AI at New York University (NYU R/AI)~\footnote{https://r-ai.co/education}.  We  conducted the workshop twice: first with professionals in news and media, and then with professionals at technology startups. Through one-on-one interviews and pre- and post-workshop surveys, we explored two research questions: (1) How effective is an educational workshop in increasing participants' algorithmic transparency literacy? and (2) Can the workshop increase  participants' willingness to advocate for algorithmic transparency in their professional 
lives?

\paragraph*{Summary of findings.} In total, 27 professionals (15 from news and media and 12 from technology startups) participated in our algorithmic transparency workshops. We divide our results into two categories: workshop-specific findings and broader findings.

\emph{Workshop findings.} Interviews with participants demonstrate that the workshops were effective in both teaching algorithmic transparency and increasing participants' willingness to advocate for it. With respect to the former, participants expressed that the workshop was particularly helpful in uncovering knowledge gaps in algorithmic transparency. Three participants noted that it made them realize ``they didn't know what they didn't know [about transparency].''

In terms of advocacy, four participants reported taking  advocacy actions in the days following the workshop. Most significantly, one participant attended an organization-wide strategy meeting on AI and spoke up on behalf of transparency, citing the workshop as a major motivator and directly applying its lessons.

Our qualitative results are also supported by pre- and post-workshop surveys, which suggest that the workshop increased participants' general understanding of algorithmic transparency as well as their willingness to advocate for it.

\emph{Broader findings.} First, we found that advocacy is not a monolith and can occur at three levels. The first is conversational, where individuals raise awareness by speaking to their colleagues about the importance of algorithmic transparency. The second is implementational, where AI engineers directly implement tools to improve transparency (\eg create data sheets~\cite{DBLP:journals/cacm/GebruMVVWDC21}, model cards~\cite{DBLP:conf/fat/MitchellWZBVHSR19}, or nutritional labels~\cite{DBLP:journals/debu/StoyanovichH19}). The third is influential, where individuals attempt to steer the overall direction of their organization's overall direction towards transparency.

Second, we found that transparency advocacy depends on the domain of use. For example, professionals in news and media are more likely to advocate for transparency but may lack the tools to act on it. In contrast, those in technology startups are more likely to have the tools and technical knowledge but lack the resources to prioritize it. Understanding these domain-specific barriers will be critical for achieving meaningful algorithmic transparency in practice.

\section{Related Work}
\label{sec:related}

\paragraph{Organizational barriers to transparency.} Organizations often forgo transparent, responsible AI practices due to misaligned incentives. This is especially true in for-profit organizations, where such practices may be perceived as barriers to increasing revenue~\cite{metcalf2019owning}. When companies do pursue responsible AI practices, it's often in response to external pressures rather than proactive, value-driven decisions by leadership~\cite{metcalf2019owning}. Employees at one large technology company reported that their day-to-day work prioritized profit-motivated tasks, such as launching products and increasing user engagement, over ethical considerations~\cite{metcalf2019owning, madaio2020co, rakova2021responsible}. In fact, the priorities of companies can, at times, be in \emph{direct tension} with responsible AI.  For example, optimizing user engagement---a common profit-driven objective---can lead to irresponsible outcomes, such as creating online radicalization pipelines~\cite{phadke2022pathways}. 

Another organizational barrier to transparency is that practitioners are at times unable to identify their companies' \emph{specific goals} with respect to broad terms like AI ethics~\cite{metcalf2019owning, raji2020closing}.  Additionally, there are human ``blind spots,''---individuals in different, disconnected teams who are unaware of responsible AI practices~\cite{holstein2019improving}. As a result, the responsibility for AI ethics often falls to motivated individuals, often referred to as ``ethics owners''~\cite{metcalf2019owning}. 

\paragraph{Regulation.} At present moment, AI regulation is insufficient for ensuring organizations are transparent about their use of algorithms. Regulation is also not a silver bullet---even among the few positive examples, omissions or loopholes exist that can be exploited. For example, the European Union AI Act establishes transparency obligations for AI systems, but differentiates the required level of transparency based on predefined AI risk categories. This approach has been criticized as a misstep, \emph{all AI systems} have the potential to pose high risk.

The majority of existing AI directives and strategies lack specificity and means of enforcement~\cite{unicri, munn2023uselessness}. As an example, consider the enacted European Union's General Data Protection Regulation (GDPR), which includes text to guarantee individuals a ``right-to-explanation,'' or a right to be given an explanation for an output of an algorithm that impacts them. However, despite being one of the most expansive and robust data protection laws to date, GDPR's right-to-explanation has yet to deliver any meaningful benefits for citizens~\cite{DBLP:conf/fat/SelbstP18,doshi2017accountability, de2022algorithmic}.

\paragraph{Tempered radicals.} Myerson coined the term ``tempered radicals'' to describe individuals who influence change within organizations slowly but steadily over time~\cite{meyerson2003tempered}. Tempered radicals prefer to make bottom-up change, rather than relying on company leadership or government regulation. There are numerous successful examples of tempered radicalism, especially in ethically motivated practices. These individuals have advanced minority representation, inclusion, and sustainability across various contexts, including companies, universities, and religious organizations~\cite{walton2013tempered, griffiths2022tempered, meyerson2007tempered, ngunjiri2012tempered, kirton2007british}.

Tempered radicals offer a natural approach to advancing responsible AI practices. Interestingly, ground-level employees already seem to bear this responsibility: interviews with researchers revealed that employees at a large tech company often feel it is \emph{their} job to represent ethical technology values~\cite{rakova2021responsible}.

\paragraph{AI education.} In recent years, there has been a growing number of initiatives teaching \emph{AI literacy}, helping citizens better understand AI at a conceptual level, including its opportunities and risks~\cite{dominguez}. While these initiatives have been primarily focused on K-12 students and emphasized the technical aspects of AI (\ie computer programming)~\cite{dominguez,williams2021train}, several promising courses have emerged that teach responsible AI to the general public~\cite{lewis2021teaching,bell2023algorithmic}. Best practices for AI education are still evolving and require a multi-disciplinary effort to incorporate the social sciences, pedagogy, and data science~\cite{lewis2021teaching}. This work intends to build up this knowledge base.

\section{Methods}
\label{sec:methods}

\renewcommand*\arraystretch{1.3}
\begin{table*}[ht!]
\small 
\centering
\caption{Modules covered in the workshop.}
\begin{tabular}{p{3.6cm}p{11cm}c}
\toprule
\textbf{Module} & \textbf{Topics} & \textbf{Time (\emph{mins})} \\
\midrule
All About Transparency & Defining algorithmic transparency, types of transparency, stakeholders and their goals & 20 \\ \hline
Transparency Tools & Transparency labels, model cards, feature importance, Shapley values, dashboards & 10 \\ \hline
The Transparency Playbook & How to disclose the use of AI, transparency for algorithms protected by IP or procured from vendors, the gold standard approach to transparency & 15 \\ \hline
Breakout Activity & Role-playing game where participants take on either the role of pro-transparency or anti-transparency managers at a fictional news and media company & 15 \\ \hline
Becoming a Transparency Advocate & Common objections to transparency (\ie ``transparency means more costs,'' ``transparency means sacrificing privacy'') and how to rebut them & 10 \\
\bottomrule
\end{tabular}
\label{tab:modules}
\end{table*}

\subsection{The Algorithmic Transparency Workshop}
\label{subsec:design}

\paragraph{Development process.} 
We designed a 2-hour workshop on algorithmic transparency, consisting of 5 modules that provide an overview of transparency, describe best practices and tools for its implementation, and outline strategies for  advocating for transparency. The workshop also includes a role-playing activity where participants act out practical barriers to implementing transparency.  Workshop materials--- including the content of the modules, the full \emph{Transparency Playbook}, and a slide-deck version of the course---are free to use and can be found on the workshop  website.

Prior to conducting the workshops, we published peer-reviewed work on a stakeholder-first approach to implementing algorithmic transparency and created a practitioner-focused playbook on the topic~\cite{bell2023algorithmic}. This work was informed by interviews with professionals across a variety of domains and backgrounds, including large technology companies, algorithmic safety audit firms, government organizations, and early-stage startups. 

We added two topics to the workshop based on practitioner input:  transparency for procured tools (common in government organizations) and balancing transparency with intellectual property considerations (common in industry). We also conducted multiple ``trial runs'' of the workshop, refining the content and the presentation based on participant feedback. For example, the \emph{Transparency Tools} module, which contains five real-world case studies of algorithmic transparency tools, emerged in response to feedback from trial run participants who suggested to add examples of how transparency is used in practice.

\paragraph{Structure and design.}
\label{para:workshop_design}
Workshop modules are summarized in Table~\ref{tab:modules}. Each module includes a lecture component, with 2-3 interactive elements, such as questions, reflections, and short discussions. For example, in the \textit{Transparency Tools} module, participants explore technical tools associated with real-world algorithmic systems, such as model cards\footnote{\url{https://www.salesforce.com/blog/model-cards-for-ai-model-transparency/}} and explainer dashboards\footnote{\url{https://titanicexplainer.herokuapp.com/multiclass}}. The module features a live demo of an explainer dashboard, followed by a discussion with the audience about the types of transparency it offers and which stakeholders it benefits. 

\paragraph{Breakout activity.}
\label{para:breakout_activity}
The moderated breakout activity aims to increase participant engagement and deepen their connection to the content, and improve participants' ability to advocate for transparency by demonstrating the tensions that emerge when organizations consider ``disclosing'' their algorithm use. For example, we aimed to highlight that some managers may object to algorithmic transparency to protect intellectual property, and to equip participants to counter these arguments.   

Participants are asked to imagine themselves as managers at a fictional social news startup (\ie \emph{HackerNews}) that had recently implemented an AI content moderation tool. Half of the participants are asked to role-play \emph{skeptical managers} (\ie the ``Devil's Advocate'' position), opposing disclosure  of the AI tool, while the other half are asked to role-play \emph{pro-transparency} managers. Participants then make and record their arguments for and against transparency for different stakeholders at their organization (\eg affected users, developers, managers, etc., as discussed in the workshop content) according to their role. An example of a completed activity can be seen in Appendix Figure~\ref{fig:breakout_activity}.

\subsection{Recruitment, Participation, and Domains of Study}

We conducted the workshop twice, each time for a different audience. The first workshop, held virtually, was attended by 15 news and media professionals. The second, conducted in person, was attended by 12 professionals working at or with technology startups. At both workshops, we administered a pre- and post-workshop surveys and conducted semi-structured follow-up interviews. 

News and media organizations and technology startups are deeply affecting (and affected by) emerging AI technologies. The release of generative AI tools like ChatGPT has significantly disrupted workflows in news and media companies, prompting existential conversations  about adaptation.  AI has had a similar impact on the startup landscape: roughly a quarter of all venture capital funding went to AI-based startups in 2023, as compared to only 11\% in 2018.\footnote{\url{https://news.crunchbase.com/ai-robotics/us-startup-funding-doubled-openai-anthropic-2023/}} 

We ran the workshops through the NYU Center for Responsible AI, in partnership with other entities at our university that work within the respective domains:  AI \& Local News at the NYC Media Lab~\footnote{\url{https://engineering.nyu.edu/research-innovation/centers/nyc-media-lab/projects/ai-local-news}} co-hosted the virtual workshop and helped recruit participants from news and media, and the NYU Tandon Future Labs~\footnote{\url{https://futurelabs.nyc/}} co-hosted the in-person workshop and helped recruit participants from their startup community.  In total, 27 domain professionals attended the workshops: 15 in the news and media workshop and 12 in the startup workshop. Nearly all participants work with AI technologies and algorithmic tools. Their job titles include Chief Digital Officer, Data Journalist, Newsroom Developer, UX Designer, Startup Co-founder, and Product Strategist. 

\paragraph{Content customization.} To improve relevance and practical applicability, we customized the content for the audience's domain. This customization manifested in two ways.  First, we tailored case studies and examples to tools and systems used in news and media or startups, respectively. For example, the news and media workshop included a discussion about the media company \emph{CNET}'s recent use of AI to generate articles on its site, many of which contained errors.\footnote{\url{https://www.theverge.com/2023/1/25/23571082/cnet-ai-written-stories-errors-corrections-red-ventures}} As part of the workshop, we discussed what went wrong and how \emph{CNET} could have benefited from a transparent AI strategy. Second, we designed the breakout activity to address a practical challenge specific to the participants' domain.

\subsection{Data Collection and Analysis}

\paragraph{Interviews.} In the days following each workshop, we conducted semi-structured interviews with 7 participants, whose domains and expertise are detailed in Table~\ref{tab:participants}. The full interview protocol is included in the Appendix. 
To encourage participants to speak candidly about their workplace experiences and their employers, we chose not to record the interviews.  Instead, we took detailed notes throughout the sessions, capturing quotes relevant to our research.

\paragraph{Pre- and post-workshop surveys.} We administered an 8-question pre-workshop survey to assess participants' baseline knowledge of algorithmic transparency and their willingness to advocate for it. Following the workshop, we administered an 18-question post-workshop survey to evaluate its impact. The survey was adapted from previous work by~\citet{lewis2021teaching}, who used a similar study design for a technical course on responsible data science. In total, 6 constructs were measured (reported in Table~\ref{tab:survey_stats}), and our choice of scale for measuring each construct was consistent with~\citet{lewis2021teaching}. Full survey details are included in the Appendix.

\paragraph{Analysis.} To analyze our qualitative data---which included both interview notes and answers to free-response questions on the post-workshop survey---we followed the six stage approach to thematic analysis described by~\citet{braun2006using}. First, we familiarized ourselves with the data by re-writing and organizing interview notes and noting initial recurrent ideas (\eg ``frequent use of AI''). Second, we generated 33 initial codes by highlighting salient interview quotes and ascribing them a code (\eg ``thinking about user needs,'' ``arbitrary thresholds for disclosure''). Interview coding was done manually using a word processor. Third and fourth, we collated the 33 codes into 6 separate themes, and evaluated their robustness over two separate working sessions. Fifth, we definitively named the themes, and, sixth, we analyzed them through the lens of our two research questions: (1) How effective is an educational workshop in increasing participants' algorithmic transparency literacy? and (2) Can the workshop increase  participants' willingness to advocate for algorithmic transparency in their professional lives, becoming \emph{transparency advocates}?

Regarding quantitative data, only 15 participants completed both the pre- and post-workshop survey (7 from news and media and 8 from technology startups). Due to  the relatively small sample size and the greater substantive value of our qualitative findings, we report survey results as descriptive statistics and forgo statistical analysis.
\section{Results}

\subsection{Thematic Analysis Findings}

\renewcommand*\arraystretch{1.3}
\begin{table*}[t]
\small 
\centering
\caption{Domain and expertise of interview participants.} 
\begin{tabular}{p{1cm}p{2.5cm}p{11cm}}
\toprule
\textbf{Alias} & \textbf{Domain} & \textbf{Expertise} \\
\midrule
P1 & Newspaper & Works for a print media company with an online presence; experience in journalism \\ \hline
P2 & Researcher & Holds a doctorate in human-computer interaction; expertise in transparency \\ \hline
P3 & Newsroom & Manages team of developers who are also journalists at popular online media company  \\ \hline
P4 & Local TV news & Works on development at syndicated local TV news network; has journalism experience \\ \hline
P5  & Startup Co-founder & CMO at an early-stage, consumer-facing, AI-based company\\ \hline
P6 & Product Manager & Worked in UI/UX design for an early-stage startup; experience in the defense industry \\ \hline
P7 & Project Manager & Works in local government using open data to improve K-12 curriculum design\\
\bottomrule
\end{tabular}
\label{tab:participants}
\end{table*}

\paragraph{Frequent use of internally developed and procured algorithmic tools.} All participants reported frequent or almost daily contact with AI in their jobs, utilizing a wide range of algorithmic tools, including generative AI, recommender systems, computer vision, and tools for carrying out A/B testing. Participants from news and media mentioned the use of both third-party and proprietary AI tools. In contrast, participants from startups relied almost exclusively on proprietary, internally developed tools.

\paragraph{Uncovering knowledge gaps.} Participants generally found the workshop useful, with each identifying different aspects as most impactful. These included learning about the different levels of transparency (P1 and P6), existing toolkits (P2 and P6), stakeholder identification (P3), and transparency tensions (P6 and P7). Several participants highlighted that the workshop's greatest strength was uncovering knowledge gaps. P2, P3, and P4 said that it helped them realize ''they didn't know what they didn't know.'' P3 reflected that, after the workshop, they realized their organization ``probably doesn't do enough disclosure and transparency.'' Interestingly, P7 offered a different perspective, stating, ``[The workshop] showed me I know a lot more than I thought I knew.''

\paragraph{Taking action.} Participants P2, P3, and P4 reported taking transparency advocacy action in the days following the workshop. P2 said they  had ``already used the [course material]'' in conversations with colleagues, and P4 noted ``I've probably had five conversations about AI transparency compared to close to zero [before the workshop].'' P3 stated that they had already begun implementing elements from the workshop, such as stakeholder identification, into their workflow. P7 said, ``I like the concept of being a transparency influencer---it shows that we can make [impacts] no matter where we are in the loop.''  

Notably, in the days immediately following the workshop, P4 stepped into the role  of a``transparency advocate'' by speaking for algorithmic transparency at an organization-wide AI strategy meeting. They described the experience: ``I was just in a TV workshop and [I asked if] we need to be disclosing and transparent [about AI], and then it got really quiet.'' But they optimistically added, ``it's definitely on the agenda now.'' 

Participants' advocacy actions appeared to be motivated, at least in part, by the workshop. Many noted that it  provided valuable resources for advocacy. As P2 explained,  ``I always would've advocated for transparency anyway ... '' but the workshop improved their potential for transparency advocacy by making them aware of different types of resources related to transparency.'' 

P3, P4, and P5 also commented on their future plans for transparency advocacy based on the resources introduced in the workshop. P3 said ``If we ever go down the road of building a model, it feels like [model cards] are something we should probably do.'' Similarly, a participant from the startup workshop, who was not interviewed but completed the post-workshop survey, wrote, ``As a co-founder at a health-tech company, I will definitely advocate for algorithmic transparency due to the benefits not only in terms of business acumen, but as a responsibility.'' 

\paragraph{Organizational challenges: resisting change} P1 and P4 reported that their organizations recently held internal meetings to discuss AI strategies and create a ``Code of Conduct'' for its use---an indication that news and media organizations are responding to the rapid proliferation of AI tools. However, both participants pointed out that this transition has not been smooth. P1 stated that discussions around the use of generative AI for creating story headlines has ``ruffled a lot of feathers,'' dividing the organization into two schools of thought: those who are \emph{pro} new AI tools, and those who are against their use and are resistant to change. Similarly, P7, who works in local government, described a comparable organizational culture where not everyone is interested in understanding AI. Reflecting on an internal AI workshop, P7 remarked,  ``[The room] was full, but relative to the amount of people in the building, it was not that full.'' They added, ``Some people are interested [in AI] but not everyone.''

\paragraph{Organizational challenges: market fundamentalism} Participants working in startups presented a different organizational challenge: while people are interested in AI, the primarily focus remains on generating revenue. When asked how often algorithmic transparency (or responsible AI) comes up as a topic of conversation at their AI-based startup, P5 said ``Not once... not with investors, not with attorneys, not with users. User’s don’t care. [Users only ask], `is it fast? Is it cool?' That’s it.'' P6 shared a similar reflection from their startup experience, saying: ``When you are in a small, bootstrapped startup, resources are tight. As a product manager, it was my job to ruthlessly prioritize [what we work on].'' They added that during rapid development cycles, ``transparency might not make the cut.'' P5 poignantly summarized this theme, stating: ``The race has begun [in AI]... [Anything not related to winning] is just not a concern... [AI start-ups] don’t care. Part of that is capitalism, part of that is behavioral.''

\paragraph{When is transparency necessary?} A surprising theme that emerged from interviews was that each participant seemed to have a personal barometer for determining \emph{when transparency is necessary.}
P1, P3, and P4 agreed that disclosing the use of AI for generating news article headlines did not seem necessary.  P6 mentioned that after the workshop, they had begun grappling with the question, ``Is more transparency always better?'' They pointed out evidence suggesting that excessive transparency about an algorithm may overwhelm end users with information~\cite{bell2022s, jacoby1974brand}. Both P5 and P6 noted that, in the startup domain, online Terms of Service and User Agreements are sometimes used to circumvent the responsibility of transparent AI practices. P6 said that the legal team often thinks, ``We can just put it in the disclaimer,'' to legally cover unethical practices. They added, ``[For users,] once they click opt-in,  it’s game over.''

Another finding within this theme was the existence of ``unwritten rules'' for transparency. For example, P1 mentioned the following guideline: ``If you are questioning whether or not you need to tell people [about AI], you need to tell people.'' Similarly, P6, drawing from their experience in the defense sector, reflected, ``There is a difference between intentionally hiding something and being intentional about what you show.''

\subsection{Pre- and Post-workshop Surveys}

\renewcommand*\arraystretch{1.3}
\begin{table*}[t]
\small 
\centering
\captionsetup{justification=centering}
\caption{Means ($\mu$) and standard deviations ($\sigma$) for pre- versus post-workshop survey responses ($n=15$)}
\label{tab:survey_stats}

\begin{tabular}{ll|cc|cc|}
 &  & \multicolumn{2}{c|}{\textbf{Pre- survey}} & \multicolumn{2}{c|}{\textbf{Post- survey}} \\ \hline
\multicolumn{1}{c|}{\textbf{Construct}} & \multicolumn{1}{c|}{\textbf{Scale}} & \multicolumn{1}{c|}{\textbf{$\mu_0$}} & \textbf{$\sigma_0$} & \multicolumn{1}{c|}{\textbf{$\mu_1$}} & \textbf{$\sigma_1$} \\ \hline
\multicolumn{1}{l|}{General Understanding of Algorithmic Transparency} & 10-point & \multicolumn{1}{c|}{5.00} & 1.80 & \multicolumn{1}{c|}{7.79} & 0.97 \\ \hline
\multicolumn{1}{l|}{Theme: Benefits and Purpose} & 3-point & \multicolumn{1}{c|}{2.00} & 0.78 & \multicolumn{1}{c|}{2.64} & 0.50 \\ \hline
\multicolumn{1}{l|}{Theme: Stakeholders} & 3-point & \multicolumn{1}{c|}{1.64} & 0.50 & \multicolumn{1}{c|}{2.57} & 0.51 \\ \hline
\multicolumn{1}{l|}{Theme: Tensions Between Goals} & 3-point & \multicolumn{1}{c|}{1.71} & 0.73 & \multicolumn{1}{c|}{2.50} & 0.52 \\ \hline
\multicolumn{1}{l|}{Willingness for Advocacy: Professional Life} & 5-point & \multicolumn{1}{c|}{4.14} & 0.77 & \multicolumn{1}{c|}{4.71} & 0.47 \\ \hline
\multicolumn{1}{l|}{Willingness for Advocacy: Personal Life} & 5-point & \multicolumn{1}{c|}{3.71} & 0.91 & \multicolumn{1}{c|}{4.43} & 0.65 \\ \hline
\end{tabular}

\end{table*}

Table~\ref{tab:survey_stats} shows the pre- and post-workshop survey results for each measured construct. Post-workshop means were higher across all constructs, with low standard deviation, suggesting that the workshop positively impacted participants' understanding of algorithmic transparency and their willingness to advocate for it. The largest improvements were observed in participants' general understanding of transparency and their awareness of transparency stakeholders. 
\section{Discussion}
\label{sec:discussion}

Overall, our findings suggest that the workshop had a positive impact both on participants' understanding of algorithmic transparency and their willingness to advocate for it. We hypothesize that this success was driven by two key factors. 

First, the workshop features a strong curriculum developed as part of a multi-year, ongoing project by the authors. As mentioned in the Methods section, the workshop was developed with practitioner feedback and has been continually improved. This iterative development underscored the importance of making the workshop freely available online so that others may use and replicate it.

Second, the workshop content was tailored to the participants' respective domains, exemplifying a stakeholder-first approach to responsible AI literacy~\cite{dominguez}. For example, P4---a professional in news and media---mentioned that their organization used AI in a manner nearly identical to the fictional scenario featured in the breakout activity of the news and media workshop. 

\subsection{Levels of Advocacy}

Promisingly, we observed several \emph{real-world} actions taken by participants following the workshops. These actions appear to have been motivated, at least in part, by the workshop. We categorize these actions into three categories: \emph{conversational, implementational, and influential}.

\paragraph{Conversational.} Three participants said that after the workshop they had more frequent conversations about algorithmic transparency with colleagues and peers. While such conversations may not \emph{directly} affect organizational change, they play a role in increasing awareness about algorithmic transparency, which in and of itself can significantly influence behaviors over time~\cite{jacobsen2011health}.

\paragraph{Implementational.} Several individuals implemented algorithmic transparency directly into their work. This type of advocacy can be characterized by narrow, immediate, and \emph{practical} changes rather than broader organizational culture shifts. As a prime example, one participant, a manager of a small team of software developers, reported integrating stakeholder identification material 
 from the workshop directly into their team's workflow. Notably, this action \emph{ did not  require organizational approval}---as a manager, they had the authority to make proactive workflow changes to promote algorithmic transparency. From our perspective, this type of advocacy is critical for driving bottom-up change within organizations, and may be consistent with the at-times clandestine actions of tempered radicals~\cite{meyerson2007tempered}.

\paragraph{Influential.} This type of advocacy is characterized by individuals taking action to affect cultural change towards algorithmic transparency within their organization. The most notable example comes from a participant who, in the days following the workshop, spoke up about algorithmic transparency at an organization-wide AI strategy meeting. They raised concerns about the organization's approach to transparency and disclosure, using arguments learned at the workshop. Interestingly, they encountered some of the same negative responses anticipated during the role-playing activity described in Methods section. Ultimately, the participant left the meeting feeling optimistic and hopeful that their company would start taking steps toward more transparent  algorithmic practices. This example highlights the potential ripple effect of the workshop: by inviting one person to think more deeply about algorithmic transparency and providing them with basic tools for advocacy, we may have contributed to a medium-sized U.S.-based media company adopting more responsible AI practices.

\subsection{The Importance of Domain-of-use}

Unexpectedly, we found vast differences in the attitudes towards algorithmic transparency in new and media vs. technology startups. For professionals in news and media, where there is an ethos of being ``champions of the truth,'' transparency and disclosure align naturally with their values. As a result, many in this domain \emph{already care about transparency}, and only need guidance on \emph{how to implement it effectively}.  On the other hand, professionals at fast-paced technology startups often \emph{cannot} afford to care about algorithmic transparency, despite possessing the technical knowledge to implement it. Although many of these professionals may care about transparency and responsible AI practices, the circumstances of AI-focused startups may prevent them from finding the time and resources to effectively act on those values. 

This finding aligns with prior researcher, which found that employees at large technology companies often express interest in value-driven work but are not given the time or space to pursue it~\cite{metcalf2019owning}. We encourage further researcher to explore whether, and how, these barriers can be overcome, noting that domain-of-use must be taken into account in such research.
\section{Conclusion, Lessons and Social Impact}
\label{sec:conlsuion}

This work outlines a promising approach to using education to affect ground-up change towards responsible AI. With this study, we hope to contribute to the broader effort to translate responsible AI practices from research settings into real-world applications, especially in high-stakes domains.

As with many studies of this nature, some findings are limited by the small sample size, posing questions about the scalability of our approach and the generalizability of our findings. This issue was further exacerbated by participant drop-off in the online workshop, which motivated us to conduct the second workshop in person. Additionally, because participation was optional, there was likely some bias toward individuals already inclined to become transparency advocates. While this may have enhanced engagement, it also highlights scalability and generalizability concerns.

We plan to continue exploring educational approaches to promote values aligned with responsible AI and encourage others to do the same. To support this effort, we have made all workshop materials used in this study publicly available online and free to use.
\section{Acknowledgements}
\label{sec:ack}

This research was supported in part by NSF Awards No. 1922658, 2326193, 2312930,
and NSF GRFP (DGE-2234660).

\begin{links}
     \link{Course website}{https://r-ai.co/transparency-playbook-course}
      \link{The algorithmic transparency playbook}{https://r-ai.co/algorithmic-transparency-playbook}
 \end{links}

\newpage 
\bibliography{aaai25}
\newpage 
\section{Appendix}
\label{sec:suppl}

On the next pages you'll find a completed breakout room activity from the one of the breakout rooms with media professionals, the pre- and post-workshop surveys, and the interview protocol.

\begin{figure*}
    \centering
    \includegraphics[width=0.8\textwidth]{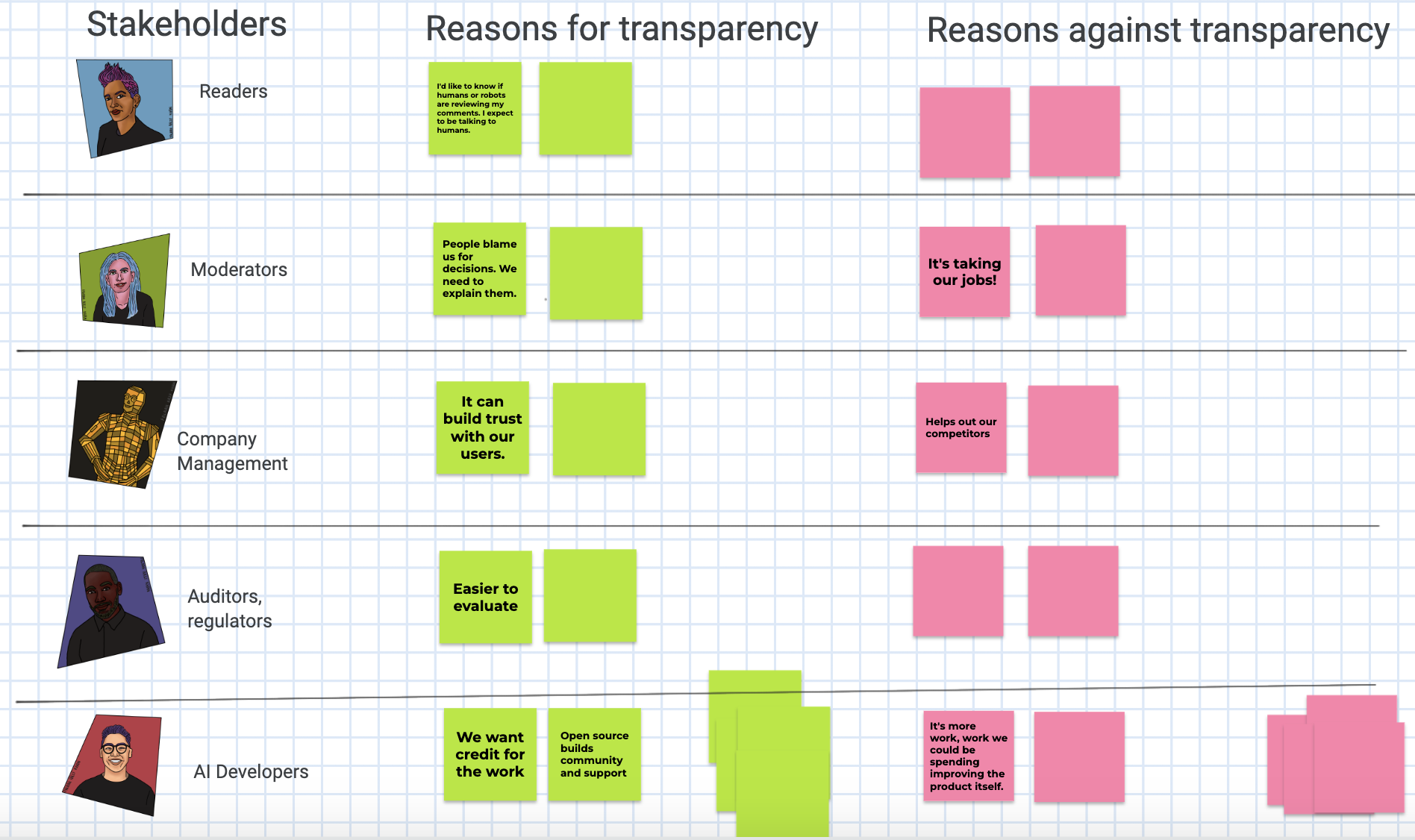}
    \caption{Completed breakout room activity from one of the four breakout rooms with media professionals. From top-to-bottom, the green cards (reasons \emph{for} transparency) read: ``I'd like to know if humans or robots are reviewing my comments. I expect to be talking to humans.'', ``People blame us for decisions. We need to explain them.'', ``It can build trust with our users.'', ``Easier to evaluate'', ``We want credit for the work'', ``Open source builds community and support.'' From top-to-bottom, the red cards (reasons \emph{against} transparency) read: ``It's taking our jobs!'', ``Helps out our competitors'', ``It's more work, work we could be spending improving the product itself.'' Illustrations by Falaah Arif Khaan.}
    \label{fig:breakout_activity}
\end{figure*}

\clearpage



\section*{Supplementary material}

# Survey questions

**Color key:**
**Pre-survey only**
**Post-survey only**
**Both pre- and post-survey**

| Theme | # | Question text | Responses |
|---|---|---|---|
| **Participant email** | - | What is your email? | [Free response] |
| **Participant Background Information** | 1 | What is your job position/title? | [Free response]<br>Not applicable |
| **General AI Sentiment** | 2a | Which two (2) words best describe how you feel about AI in the world right now? | Excited<br>Hopeful<br>Curious<br>Ambivalent<br>Concerned<br>Nervous<br>Scared as hell<br>Uninformed<br>Other |
| | 2b | Please tell us why | [Free response] |
| | 3 | Do you think AI will make our lives better? | Yes<br>Somewhat<br>No<br>Not sure |
| | 4 | Do you consider the responsible use of AI to be important for your work? | Yes<br>Somewhat<br>No<br>Not sure<br>Not applicable |
| **Transparency Knowledge A** | 5 | How do you rate your current general understanding of concepts in algorithmic transparency? | 1 to 10 from "What is algorithmic transparency?" to "I'm an expert in algorithmic transparency" |
| | 6 | How do you rate your current understanding of these specific workshop themes as they relate to the transparency of algorithmic decision systems?<br><br>(1) Benefits and purpose of transparency; (2) Transparency stakeholders and their goals; (3) Tensions between transparency and other | For each of the 3 themes:<br>Fuzzy understanding<br>Moderate understanding<br>Great understanding |

| | | | |
|---|---|---|---|
| | | goals | |
| **Transparency advocacy** | 7a | How likely are you to advocate for algorithmic transparency in your work? | 1 to 5 from "Very unlikely" to "Very likely"<br>Not applicable |
| | 7b | Please tell us why | [Free response] |
| | 8a | How likely are you to advocate for algorithmic transparency in your personal life? | 1 to 5 from "Very unlikely" to "Very likely"<br>Not applicable |
| | 8b | Please tell us why | [Free response] |
| **Transparency Knowledge B** | 9 | To what extent was your understanding of algorithmic transparency concepts and techniques improved by the workshop? | 1 to 5 from "Not at all" to "Substantially improved" |
| | 10 | Do you feel that you'd be able to give a brief overview of algorithmic transparency to a friend? | Yes<br>No<br>Not sure |
| | 11a | How motivated are you to apply what you learned during the workshop in your work? | 1 to 5 from "Not at all" to "Substantially"<br>Not applicable |
| | 11b | Please tell us why | [Free response] |
| **Learning and Session Feedback** | 12 | Did the workshop cover the content you expected it would cover? | Yes<br>Somewhat<br>No<br>I had no expectations |
| | 13 | The workshop provided enough background on algorithmic transparency to position you to engage constructively with the material. | 1 to 5 from "Strongly disagree" to "Strongly agree" |
| | 14 | The workshop makes me think more critically and effectively about algorithmic transparency. | 1 to 5 from "Strongly disagree" to "Strongly agree" |

| | 15 | The workshop supported your comprehension of algorithmic transparency concepts and techniques. | 1 to 5 from "Strongly disagree" to "Strongly agree" |
|---|---|---|---|
| | 16 | The workshop structure was clear. | 1 to 5 from "Strongly disagree" to "Strongly agree" |
| | 17 | The content was presented clearly. | 1 to 5 from "Strongly disagree" to "Strongly agree" |
| | 18 | The length of the workshop was appropriate. | Way too long<br>A bit too long<br>Just right<br>A bit too short<br>Way too short |
| | 19 | How likely are you to recommend this course to a friend, family member or colleague? | 1 to 5 from "Very unlikely" to "Very likely"<br>Not applicable |

**Interview Protocol**

- What is your job position/title?
- How often do you use AI in your professional work?
    - Probe: if you do not explicitly use AI in your work, are there ways in which you indirectly interact or interface with AI?
- How well do you feel you currently understand algorithmic transparency?
    - Probe: where do you think there may be gaps in your understanding?
    - Probe: what themes of algorithmic transparency do you best understand?
- How do you feel the *Workshop on Algorithmic Transparency* helped your understanding of algorithmic transparency (if at all)?
    - Probe: what do you feel was the greatest strength of the course?
    - Probe: what do you feel was the greatest weakness of the course?
- In your opinion, why do organizations (possibly even the organization where you work) forgo algorithmic transparency?
    - Probe: have you personally come across any reasons in your professional work?
- Have you ever advocated for transparent AI practices in your professional work?
    - Probe: what was your advocacy experience like, i.e. were your ideas well-received or dismissed by peers and leadership?
    - Probe: *would you consider* advocating for transparent AI practices in your work?
- How do you feel the *Workshop on Algorithmic Transparency* helped your understanding of algorithmic transparency advocacy (if at all)?
    - (If applicable) what stops you from advocating for transparency in AI?
    - (If applicable) what motivates you to advocate for transparency in AI?
- How often do you discuss algorithmic transparency in your personal life?
    - Probe: what kind of response do you hear/reactions do you see when speaking about algorithmic transparency?



\end{document}